\documentclass[aps,prl,twocolumn,superscriptaddress,floatfix,showpacs]{revtex4}

\usepackage{graphicx}
\usepackage{amsmath}
\usepackage{color}

\def\cd{c^{\dagger}}
\def\eps{\varepsilon}
\def\s{\sigma}
\def\w{i\omega_n}
\def\bk{\mathbf{k}}

\def\etal{\textit{et al.}}

\begin{document}
\title{Study of the specific heat for the binary alloy in the CPA+DMFT method.}

\author{Alexander I. Poteryaev}
\affiliation{M.N. Miheev Institute of Metal Physics of Ural Branch of Russian Academy of Sciences, S. Kovalevskaya str. 18, Ekaterinburg 620137, Russia}
\author{Sergey L. Skornyakov}
\affiliation{M.N. Miheev Institute of Metal Physics of Ural Branch of Russian Academy of Sciences, S. Kovalevskaya str. 18, Ekaterinburg 620137, Russia}
\affiliation{Ural Federal University, 620002 Ekaterinburg, Russia}
\author{Alexander S. Belozerov}
\affiliation{M.N. Miheev Institute of Metal Physics of Ural Branch of Russian Academy of Sciences, S. Kovalevskaya str. 18, Ekaterinburg 620137, Russia}
\affiliation{Ural Federal University, 620002 Ekaterinburg, Russia}
\author{Vladimir I. Anisimov}
\affiliation{M.N. Miheev Institute of Metal Physics of Ural Branch of Russian Academy of Sciences, S. Kovalevskaya str. 18, Ekaterinburg 620137, Russia}
\affiliation{Ural Federal University, 620002 Ekaterinburg, Russia}


\begin{abstract}
The thermodynamic properties of strongly correlated system with binary type of disorder are investigated
using the combination of the coherent potential approximation and dynamical mean-field theory.
The specific heat has a peak at small temperatures for the concentrations close to the filling of system.
This peak is associated with the local moment formation due to Coulomb interaction.
The linear coefficient to the specific heat is divergent and the system stays in the non-Fermi-liquid regime. 
\end{abstract}


\maketitle

\section{Introduction}

The investigation of the properties of real materials from first principles is one of the challenging
problems in condensed matter physics. Over the last half of century the methods within density functional theory
framework recommended themselves as very successful at describing the wide band compounds. 
Concurrently the coherent potential approximation~\cite{Elliott74,Ruban2008,Rowlands2009} (CPA) 
was developed to treat the effects of disorder in alloys and real materials that are far from perfect crystal stoichiometry.
The physics of transition metal alloys and steels are defined by the doping components and hence
it is highly desirable to have a method that can treat on equal footing disorder and
partially filled strongly interacting $d$ states of transition metals.
The widely accepted solution for problem of interacting electrons is dynamical
mean field theory~\cite{Georges1996} (DMFT) that has a common feature with CPA -- the effective medium interpretation
of the system of interest. 

On the model level the CPA+DMFT method was introduced by Jani\v{s} \etal~\cite{Janis1992,Janis1993}
almost two decades ago where they combined coherent potential approximation
with dynamical mean field theory to treat problem of disorder and interacting electrons simultaneously.
Later on they studied in details the thermodynamic properties for the Anderson-Hubbard model 
with different type of disorder distributions for half-filling and constructed magnetic phase diagram~\cite{Ulmke1995}.
In the same years, Bhatt and Fisher~\cite{Bhatt1992} and Dobrosavljevi\'{c} and Kotliar~\cite{Dobrosavljevic1994},
who also studied half-filled system,
found non Fermi liquid behavior of interacting systems with different type of disorder at low temperatures
that manifested by divergent magnetic susceptibility and linear coefficient at specific heat, $\gamma$, 
induced by local moments.
Laad \etal~\cite{Laad2001} studied the spectral properties and observed that at small values 
of disorder strength the renormalized Fermi liquid behavior is retaining while increase
of disorder results in the incoherent metallic and than insulating state.
Byczuk and co-authors in a number of papers~\cite{Byczuk2003,Byczuk2004,Byczuk2005,Byczuk2010} 
studied different properties of the Anderson-Hubbard model.
They found that disorder can enhance the Curie temperature with respect to a perfect system~\cite{Byczuk2003,Byczuk2005}
and for large values of disorder strength the metal-insulator transition can occur at non-integer filling equal to 
concentration~\cite{Byczuk2003,Byczuk2004}.
In case of full diagonal disorder Lombardo \etal~\cite{Lombardo2008} found 
that transition from a band insulator to a metallic state and further to a Mott-Hubbard insulator 
can be driven by the correlation strength of only one of the alloy constituents. 

The specific heat is one of the easy-to-measure and, at the same time, important quantity
that gives an access to the many thermodynamic properties of real materials.
Therefore, the understanding of its behavior upon doping, temperature, pressure or other
external parameters is of high importance for physics of alloys.
In the present paper we investigate the specific heat in the disordered Hubbard model
with the local binary disorder as a function of temperature and doping.
This will allow one for better explanation of the steel's and alloy's properties.

\section{Method}

The Hubbard-type Hamiltonian of the binary alloy system $A_xB_{1-x}$ 
with substitutional type of disorder can be written as
\begin{equation}
  H = - \sum_{\langle ij \rangle, \s} t_{ij} \cd_{i\s} c_{j\s} + \sum_{i\s} ( \eps_{i\s} - \mu ) n_{i\s}
      + \sum_i U_i n_i^{\uparrow} n_i^{\downarrow},
  \label{eq:hamilt}
\end{equation}
where $\cd_{i\s}$ ($c_{i\s}$) is a creation (annihilation) operator and $n_{i\s}= \cd_{i\s}c_{i\s}$
($i$, $j$, $\s$ are the site and spin indexes).
Therefore, the first term is responsible for the transfer of the electrons from one site to other
with the hopping amplitude $t_{ij}$. 
The second term contains $\eps_{i\s}$ and $\mu$ that are one site and chemical potentials.
The former is equal to either $\eps_A$ or $\eps_B$ (case of so-called diagonal disorder)
with the probability $x_A$ or $x_B$ ($x_A+x_B=1$). 
The last term of this Hamiltonian is an interacting contribution,
like in conventional Hubbard model, but different for the different atomic species,
$U_A$ or $U_B$, depending on atom remaining on site $i$.
$x_A$ ($x_B$) is a concentration of the atoms of type $A$ ($B$).

The complexity of the above Hubbard-type Hamiltonian is high enough to be solved directly but
one can follow the effective medium ideology applied in coherent potential approximation
for disordered systems and later in dynamical mean-field theory to treat correlated problem.
In this case an action for the effective medium problem is
\begin{equation}
  S_{med} = - \int_0^{\beta} d \tau \int_0^{\beta} d \tau' \sum_{\bk\sigma} \cd_{\bk\sigma}(\tau) 
            G^{-1}_{med}(\bk,\tau-\tau') c_{\bk\sigma}(\tau')
  \label{eq:s_med}
\end{equation}
with Green function of effective medium
\begin{equation}
  G^{-1}_{med}(\bk,\w) = \w + \mu - H(\bk) - \Sigma(\w),
\end{equation}
where $\beta=1/T$ is an inverse temperature, 
$\w=(2n-1)\pi/\beta$ are fermionic Matsubara frequencies and 
$H(\bk)$ is a Fourier transform of the kinetic term of the Hamiltonian.
The self-energy, $\Sigma(\w)$, is a pure local quantity that contains 
the information about disorder and local on-site interaction and has to be
determined self-consistently.

One can easily write an impurity action embedded in the effective medium
by subtracting the local self-energy on site and adding back the exact interaction
\begin{align}     \nonumber
  S_{imp}^{\alpha} & = S_{med} - 
     \int_0^{\beta} d \tau \int_0^{\beta} d \tau' \sum_{\sigma} 
             \cd_{\alpha\sigma}(\tau) \Sigma(\tau-\tau') c_{\alpha\sigma}(\tau')       \\  
           & + U_{\alpha} \int_0^{\beta} d \tau n_{\alpha}^{\uparrow}(\tau) n_{\alpha}^{\downarrow}(\tau)
             + \eps_{\alpha} \int_0^{\beta} d \tau \sum_{\sigma} n_{\alpha\s}(\tau)
\end{align}
with $\alpha=\{A,B\}$. The sites distinct from $\alpha$ can be integrated out exactly
because they enter quadratically in action and it now becomes
\begin{align}     \nonumber
  S_{imp}^{\alpha} & =  - 
     \int_0^{\beta} d \tau \int_0^{\beta} d \tau' \sum_{\sigma} 
             \cd_{\alpha\sigma}(\tau) \mathcal{G}^{-1}_0(\tau-\tau') c_{\alpha\sigma}(\tau')       \\  
           & + U_{\alpha} \int_0^{\beta} d \tau n_{\alpha}^{\uparrow}(\tau) n_{\alpha}^{\downarrow}(\tau)
             + \eps_{\alpha} \int_0^{\beta} d \tau \sum_{\sigma} n_{\alpha\s}(\tau)
\end{align}
with bath Green function
\begin{equation}
  \mathcal{G}^{-1}_0(\w) = G^{-1}_{med}(\w) + \Sigma(\w).
\end{equation}
The self-consistency condition requires that the impurity Green function embedded in effective medium
has to coincide with the local medium Green function and in case of alloy it can be written as 
properly weighted sum of Green functions for different atomic species
\begin{equation}
  \sum_{\alpha=A,B} x_{\alpha} G_{\alpha}(\w)  = \sum_{\bk} G_{med} (\bk,\w)
\end{equation}
where 
\begin{equation}
  G_{\alpha}(\tau-\tau') \equiv - \big\langle T c(\tau) \cd(\tau') \big\rangle_{S_{imp}^{\alpha}}.
\end{equation}
One should note that the effective impurity Green functions can be evaluated with any suitable technique.
Above set of equations has to be iterated until self-consistency with respect to self-energy is reached.

The measurable quantities in presence of disorder are evaluated in the usual CPA way and are regarded as
weighted sum of the local values
\begin{equation}
  \langle \mathcal{A} \rangle = \sum_{\alpha} x_{\alpha} \langle \mathcal{A}_{\alpha} \rangle,
\end{equation}
where $\alpha$ subscribe means that this is an average of operator
$\mathcal{A}$ for the certain type of atom in effective medium.
Therefore the internal energy of the system can be written as
\begin{align}   \nonumber
  E & \equiv \langle H \rangle = \frac{1}{\beta} \sum_{n,\bk} H(\bk) G_{med}(\bk,\w) e^{\w0^+}     \\
    & + \sum_{\s}\sum_{\alpha=A,B} x_{\alpha} \eps_{\alpha\s} n_{\alpha\s}
      + \sum_{\alpha=A,B} x_{\alpha} U_{\alpha} \langle n_{\alpha}^{\uparrow} n_{\alpha}^{\downarrow} \rangle
  \label{eq:total_energy}
\end{align}
where $\langle n^{\uparrow} n^{\downarrow} \rangle$ correlator can be calculated within segment version
of hybridization expansion continuous-time quantum Monte Carlo method~\cite{Werner2006} (CT-QMC) at no additional cost.

\section{Results and discussions}

We carried out the calculations of the specific heat for the binary alloy using the CT-QMC method~\cite{Werner2006}.
In this case the the interaction contribution to the internal energy can be easily calculated with high precision
and what is also important in our study that the low temperatures can be accessed.
The specific heat was evaluated as $C_v = \partial E(T) / \partial T$.
The conventional Bethe lattice is explored with the half-bandwidth, $D$=1, and in following 
all energy quantities will be expressed in units of half-bandwidth.
The total occupation was fixed to $n$=0.3 and the Coulomb interactions are equal, $U_A$=$U_B$=2,
in order to deal with only disorder due to local potentials of the different species:
$\eps_A$=-0.5, $\eps_B$=0.5. 

\begin{figure}[tbh]
  \centering                 
  \includegraphics[clip=true,width=0.45\textwidth,angle=0]{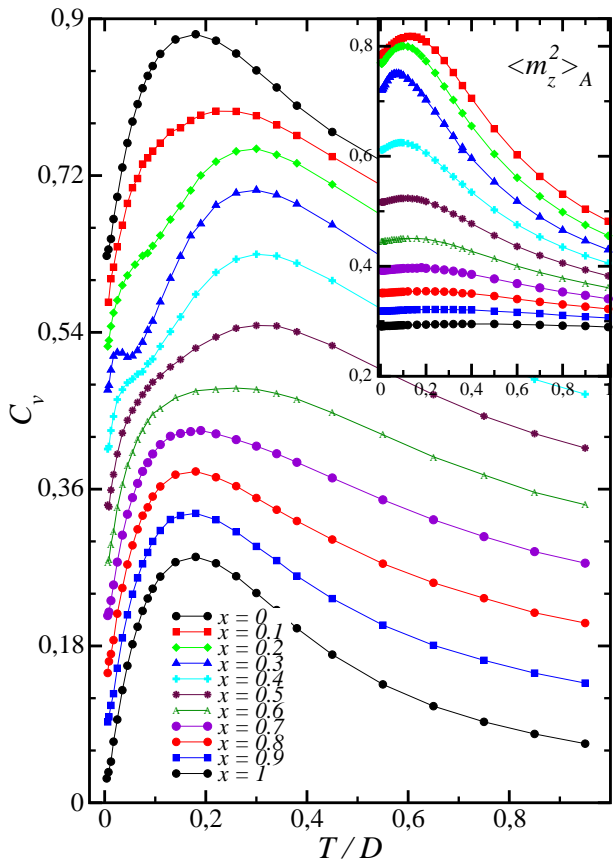}
  \caption{Specific heat versus temperature for various alloy concentrations (see colorcoding on the figure).
           Total occupation is $n$=0.3, Coulomb interactions are $U_A$=$U_B$=2, 
           local potentials are $\eps_A$=$-\eps_B$=-0.5.
           Inset shows the temperature dependence of the instant squared magnetic moment for the $A$ constituent,
           $\langle m_z^2 \rangle_A$. The curves are shifted up to make them more visible.
           }
  \label{fig:cv}
\end{figure}
The results of the calculations for various alloy concentrations, $x$, are presented on the Fig.~\ref{fig:cv}
(hereafter the concentration $x \equiv x_A$ will be used for simplicity keeping in mind that $x_A+x_B=1$).
One can clearly see common features of the specific heat at different concentrations.
At small temperature $C_v$ has small value that increases with temperature and 
at about 0.2-0.3 it has a large peak that is connected with the thermal activation
of incoherent states~\cite{Rozenberg1994}. After the specific heat is decreasing with temperature and
has an universal behavior at temperatures that are much larger than other energy scales in the model (not shown).
One should note that the position of this large peak in $C_v$ depends on the concentration. It lies at 0.18 in
pure ($x$=0) Bethe lattice and shifts to higher temperature 0.3 at concentration $x$=0.3,
afterwards the peak position returns to the value of pure position.

\begin{figure}[tbh]
  \centering                 
  \includegraphics[clip=true,width=0.33\textwidth,angle=270]{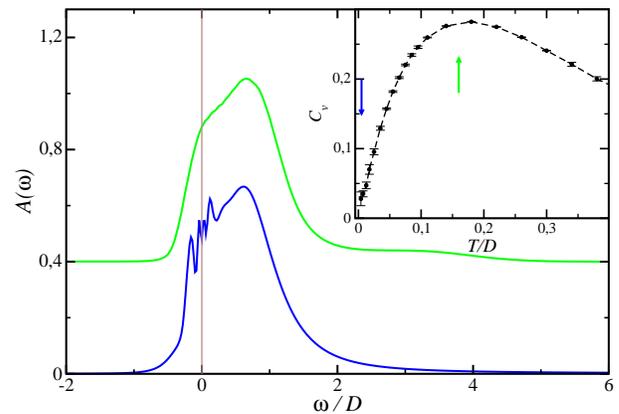}
  \caption{Evolution of spectral function with temperature for pure Bethe lattice ($x$=0).
           The spectral functions are shown for $T$=0.005 and 0.16 by blue and green colors, respectively.
           In inset the specific heat with errorbars for $x$=0 is presented. 
           Colored arrows indicate the temperatures at which spectral functions were evaluated.}
  \label{fig:dos_x=0}
\end{figure}
The most interesting behavior of the specific heat is observed at very small temperatures, $T$$<$0.1.
This can be understood from the behavior of the instant squared magnetic moment for the $A$ constituent,
$\langle m_z^2 \rangle_A$, shown in the inset of Fig.~\ref{fig:cv}.
At concentrations $x \approx n$, it has a large value at the smallest 
calculated temperatures. Than $\langle m_z^2 \rangle_A$ increases with the temperature and has a peak at the same positions
where the peak in the specific heat is located (for the corresponding concentrations).
After the instant squared magnetic moment decreases rapidly with the temperature raise.
This strong temperature dependence of the $\langle m_z^2 \rangle_A$ at these concentrations
is reflected in the large temperature dependence of the corresponding energy term 
in the eq.~(\ref{eq:total_energy}), and thus, in the specific heat.
With increasing of the concentration the magnitude of the instant squared magnetic moment for
the $A$ constituent is smaller and its temperature dependence is very weak that comes from 
the overdoped regime. Therefore the contribution to the specific heat due to correlation effects
is not strong anymore. One should note here that other terms
in total energy (eq.~\ref{eq:total_energy}) are of small temperature dependence and its
contributions to the specific heat are negligible or about the same value for the different concentrations.

\begin{figure}[tbh]
  \centering                 
  \includegraphics[clip=true,width=0.33\textwidth,angle=270]{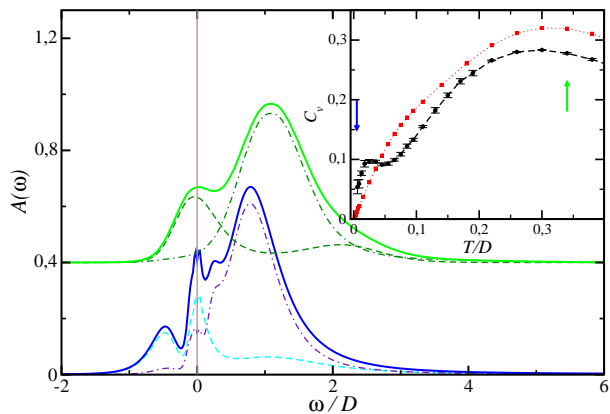}
  \caption{Evolution of spectral functions with temperature for binary alloy with $x$=0.3.
           The spectral functions are shown for $T$=0.005 and 0.36 by blue and green colors, respectively.
           The weighted partial contributions of site $A$ and $B$ are shown by dashed and dot-dashed colors.
           In inset the specific heat for $x$=0.3 is presented for the interacting and non-interacting systems by black
           dots and red squares. Colored arrows indicate the temperatures at which spectral functions were evaluated.}
  \label{fig:dos_x=0.3}
\end{figure}

The behavior of the specific heat can be also analyzed from the spectral properties view.
For the pure Hubbard model specific heat has no any features at small temperatures (see inset of the Fig.~\ref{fig:dos_x=0})
like in weakly correlated system~\cite{Rozenberg1994}.
Indeed, at these values of interaction parameters and occupation, $n=0.3$, 
the system is overdoped that one can see from the spectral function presented on the Fig.~\ref{fig:dos_x=0}.
With increasing of the concentration, $x$, the small peak starts to grow and it has maximum 
at the concentration, $x$=0.3, than this peak becomes smaller and totally disappears at $x$=0.7
(see Fig.~\ref{fig:cv}).
From the analysis of the spectral functions and its temperature dependence for $x$=0.3 (see Fig.~\ref{fig:dos_x=0.3})
one see that at $T$=0.005 the spectral function for the $A$ constituent
is almost ``half-filled'' and has a quasiparticle peak. With increase of temperature the quasiparticle peak and 
low Hubbard band are washed out in one structure and the small feature in the specific heat vanishes.
The specific heat for the non-interacting binary alloy with the same local potentials
does not show any features at $T$$<$0.1. Hence, we can clearly associate 
this peak in specific heat with the local moments formed by on-site Coulomb interaction.

It should be noted that binary alloy with larger values of the Coulomb parameters and
disorder strength ($\eps_B - \eps_A$) is insulator for the concentration equal to filling
of the system, $x=n$~\cite{Byczuk2004}. This transition from weakly to strongly correlated state can be explained as follow.
Let suppose that there is a binary alloy with large disorder strength, $\eps_B - \eps_A > D$, and interaction
parameters that are also comparable with the bandwidth, $U \gtrsim D$.
The filling, as in our case, is $n$=0.3 and is different from integer. 
When the concentration is equal to one we have a pure Hubbard model, all $A$ sites are occupied by 0.3 electrons 
and the system is in weakly correlated regime due to overdoping.
At decreasing the concentration the number of $A$ sites is decreased and the occupation, $n_A$, is increased
because the local potential of $B$ sites is higher in energy and they are almost empty.
When the concentration becomes equal to the (per site) filling of the system the $A$ sites are close
to half-filling, and thus, they are in strongly correlated regime. 
Therefore, the system under investigation is on the verge
of the metal-insulator transition and local moments degrees of freedom are of high importance.

\begin{figure}[tbh]
  \centering                 
  \includegraphics[clip=true,width=0.33\textwidth,angle=270]{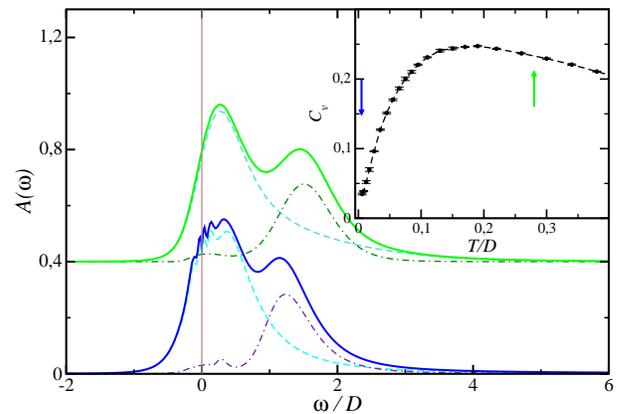}
  \caption{Evolution of the spectral functions with temperature for binary alloy with $x$=0.7.
           The spectral functions are shown for $T$=0.005 and 0.28 by blue and green colors, respectively.
           The partial contributions of site $A$ and $B$ are shown by dashed and dot-dashed colors.
           In inset the specific heat for $x$=0.7 is presented. 
           Colored arrows indicate the temperatures at which spectral functions were evaluated.}
  \label{fig:dos_x=0.7}
\end{figure}

The spectral functions for $x$=0.7 are shown on the Fig.~\ref{fig:dos_x=0.7}.
Neither spectral functions for the $A$ sites nor for the $B$ sites show correlated features
at small temperature, and thus, the specific heat does not have the peak. 
The raise of temperature leads to smoothing of the spectral functions without qualitative changes.
\begin{figure}[tbh]
  \centering                 
  \includegraphics[clip=true,width=0.33\textwidth,angle=270]{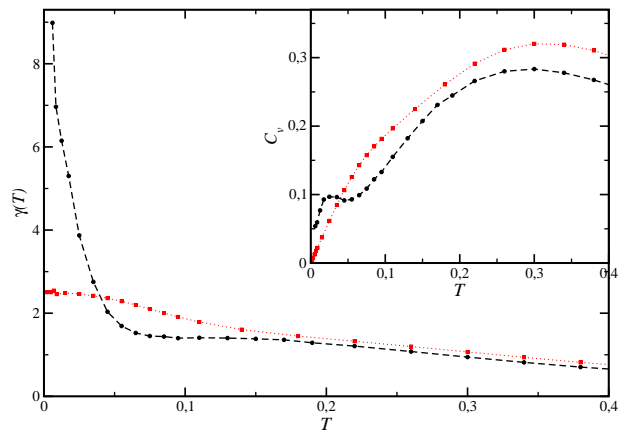}
  \caption{Temperature dependence of $\gamma(T)$=$C_v(T)/T$ for interacting (black) and non-interacting (red)
           binary alloy with concentration, $x$=0.3. Inset shows the corresponding specific heats with the same colors.}
  \label{fig:gamma_x=0.3}
\end{figure}
The temperature dependence of the linear coefficient to the specific heat, $\gamma(T)$=$C_v(T)/T$, is plotted
on the Fig.~\ref{fig:gamma_x=0.3} for the concentration $x$=0.3.
One can clearly see that in non-interacting case (red squares) the linear coefficient is constant at small
temperatures, and hence, the system is in Fermi liquid regime even in the presence of the relatively
large disorder, $\eps_B - \eps_A$=1. In interacting case the situation is different,
$\gamma$ is divergent at small temperatures~\cite{lowT} and therefore the system is in non-Fermi-liquid state.
This result is consistent with the results of Dobrosavljevi\'{c} and Kotliar
who found similar divergent behavior of the linear coefficient in the model with random hoppings and
associated it also with the local moments formation~\cite{Dobrosavljevic1993}.

To summarize, we have studied the temperature and concentration dependence of the electronic specific heat in the binary
alloy with Coulomb correlations. At very low temperatures the specific heat has the small peak which
exists at the concentrations close to the per site filling of the system, $x \approx n$.
The local magnetic moments are formed and as a consequence the peak in the specific heat appears
due to additional degrees of freedom.
The weakly correlated binary alloy can be driven to the strongly correlated regime with change of 
concentration, and hence, the magnetic properties can be tuned by concentration~\cite{Byczuk2005}. 
Additional analysis of the linear coefficient to specific heat shows non-Fermi-liquid behavior
of correlated binary alloy.

The authors thank to Rubtsov and Lichtenstein for valuable and fruitful discussions.
This study was supported by the grant of the Russian Scientific Foundation (project no. 14-22-00004).

\bibliographystyle{apsrev}

\begin{thebibliography}{99}
\bibitem{lowT} We did not found linear behavior of specific heat coefficient, $\gamma$,
          at the smallest accessible to us temperatures.
\end{thebibliography}

\end{document}